%% file: tof.tex
\documentclass[aps,prd,twocolumn,showpacs,superscriptaddress,preprintnumbers]{revtex4}  
\usepackage{graphicx}  
\usepackage{dcolumn}   
\usepackage{bm}        
\usepackage{amssymb}   
\usepackage{xspace}
\usepackage{multirow}
\usepackage{color}
\usepackage{units}
\usepackage{endnotes}

\usepackage{tikz}

\newcounter{rowcounter}
\addtocounter{rowcounter}{-1} 

\begin{document}


\title{Precision measurement of the speed of propagation of neutrinos
  using the MINOS detectors}

\input{TOF-Nov14-authors.tex} 
\date{\today}
\preprint{FERMILAB-PUB-15-289-ND}
\preprint{arXiv:1507.04328 [hep-ex]}

\begin{abstract}
  We report a two-detector measurement of the propagation speed of
  neutrinos over a baseline of \unit[734]{km}.  The measurement was made
  with the NuMI beam at Fermilab between the near and far MINOS
  detectors. The fractional difference between the neutrino speed and
  the speed of light is determined to be $(v/c-1) = (1.0 \pm 1.1) \times
  10^{-6}$, consistent with relativistic neutrinos.
\end{abstract}

\maketitle
 
\section{Introduction}

A cornerstone of the theory of relativity is that there is a single
limiting speed, the speed of light in a vacuum $c$, which cannot be
exceeded. Observations of neutrinos from
SN1987A\,\cite{Hirata:1987hu,Bionta:1987qt,Alekseev:1987ej} and
accelerator experiments\,\cite{Kalbfleisch:1979rm,Adamson:2007zzb} have
set limits on the difference between the speed of neutrino propagation
and that of light, all consistent with $v=c$. In September 2011, the
OPERA experiment reported a measurement\,\cite{Adam:2011}, in striking
conflict with both theory and experiment, which has since been revised
to resolve the inconsistency\,\cite{Adam:2012}. The initial OPERA news
motivated a number of further
measurements\,\cite{Antonello:2012hg,Agafonova:2012rh,Antonello:2012be,Adam:2012pk,AlvarezSanchez:2012wg}.
We report here a new precision measurement of the speed of neutrinos
using the NuMI muon neutrino beam at the Fermi National Accelerator
Laboratory (Fermilab)\,\cite{Adamson:2015dkw} and the two MINOS
detectors\,\cite{ref:minosnim}, using a significantly upgraded time
synchronization system and an exposure of $0.8\times10^{20}$ protons on
target in March and April of 2012.  This measurement has similar
precision to the CERN to LNGS measurements referenced above, but with
two significant differences in addition to being made in a different lab
with different dominant uncertainties.  First, the NuMI beam has a mean
neutrino energy of \unit[2.8]{GeV}, almost an order of magnitude lower
than the \unit[17]{GeV} CNGS beam.  Second, the measurement presented
here measures the neutrino time-of-flight using two neutrino detectors,
rather than using proton bunches in the accelerator to obtain the start
time and neutrinos for the stop time.  This neutrino time-of-flight
measurement is the most precise ever, although the velocity precision is
more limited by distance uncertainties.

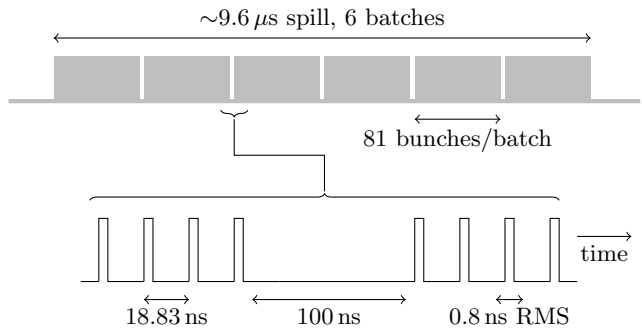
\begin{figure}
\centering
\begin{tikzpicture}[scale=0.12]
\usetikzlibrary{snakes}

\draw[gray!50,ultra thick] (0,30) -- (70,30);
\fill[gray!50] (5,30) rectangle (14.5,35);
\fill[gray!50] (15,30) rectangle (24.5,35);
\fill[gray!50] (25,30) rectangle (34.5,35);
\fill[gray!50] (35,30) rectangle (44.5,35);
\fill[gray!50] (45,30) rectangle (54.5,35);
\fill[gray!50] (55,30) rectangle (64.5,35);

\draw[snake=brace,mirror snake] (23.5,29) -- (26.5,29);
\draw (25,28) |- ( 27,24) -| (35,20);
\draw[snake=brace] (9,19) -- (61,19);

\foreach \x in {10,15,20,25,45,50,55}
 \draw[xshift=\x cm] (0,10) -- (0,17) -- (1,17) -- (1,10) -- (5,10);
\draw (30,10) -- (45,10);
\draw[xshift=60 cm] (0,10) -- (0,17) -- (1,17) -- (1,10) -- (3,10);
\draw (8,10) -- (10,10);
\draw[<->] (15,8) -- node[below] {\unit[18.83]{ns}} (20,8);
\draw[<->] (27,8) -- node[below] {\unit[100]{ns}} (44,8);
\draw[<->] (54,8) -- node[below] {\unit[0.8]{ns} RMS} (57,8);
\draw[->] (63,15) -- node[below] {time} (69,15);
\draw[<->] (5,37) -- node[above] {\unit[$\sim$9.6]{$\mu$s} spill, 6 batches} (64.5,37);
\draw[<->] (45,28) -- node[below] {81 bunches/batch} (54.5,28);

\end{tikzpicture}
\caption{Schematic of the spill structure showing the 6 batches (top),
  the gap between two of the batches (bottom), and the bunch gap and width.\label{fig:TBeam}
}
\end{figure}

The neutrino velocity measurement is conceptually straightforward,
consisting of a measurement of the distance between the two detectors
and the time it takes for a neutrino to pass between them. We can never
observe the same neutrino in both detectors since the process of
detection is destructive.  We address this issue by making two separate
measurements with respect to the same time reference.  Specifically, we
first make a measurement of the time of arrival of a bunch of neutrinos
in the MINOS Near Detector (ND) referenced to a time marker derived from
the proton beam current near the neutrino production target and then of
that bunch's arrival at the MINOS Far Detector (FD), also with respect
to the proton time reference.  Subtraction of these two measurements,
corrected for various offsets from the detection process, gives the
time-of-flight of the neutrinos over the distance between the two
detectors.  Care is taken so that when the subtraction of the two times
is made, the major part of the uncertainties in the detection cancels,
leaving a high precision determination of the time the neutrinos took to
travel from the near detector to the far detector.

The measurement uses a system of time transfer between the beam
current measurement and each detector with sets of periodically
calibrated Global Positioning System (GPS) receivers and atomic clocks. 
Two Way Satellite Time and Transfer (TWSTT) is
also used as an independent technique to calibrate the time
offsets. The combination of these techniques allowed reliable
estimates of the time synchronization errors, yielding a very robust
measurement that has the smallest error for the flight time of
neutrinos ever achieved.

\section{Beam and Detectors}

The neutrino beam\,\cite{Adamson:2015dkw} is produced at Fermilab by
\unit[120]{GeV/$c$} protons striking a graphite target. The resulting
positively charged pions and kaons are focused by pulsed magnetic
horns and then allowed to decay in a \unit[675]{m} long
helium-filled volume, producing a $\nu_\mu$ dominated neutrino beam with
a peak in the event energy spectrum at about \unit[3]{GeV} and a tail at higher
energies\,\cite{ref:combinedNuMu}. The time structure of the beam is
illustrated in Fig.~\ref{fig:TBeam}. During acceleration, the protons
are grouped into six batches, each approximately \unit[1.6]{$\mu$s}
long and separated by about \unit[100]{ns}. Each batch consists of
81~bunches which are \unit[0.8]{ns} RMS wide (\unit[3.5]{ns} full width
at base), spaced at \unit[18.83]{ns} intervals, resulting from the Main
Injector's \unit[53.103480]{MHz} synchrotron acceleration.

\subsection{The MINOS Detectors}

The two MINOS detectors\,\cite{ref:minosnim} are steel and scintillator
tracking calorimeters with toroidal magnetic fields averaging
\unit[1.3]{T} in the steel. Each detector consists of \unit[2.54]{cm}
thick steel plates interleaved with \unit[1]{cm} thick plastic
scintillator planes. The scintillator planes are composed of
\unit[4.1]{cm} wide strips.  Scintillation light is read out by
multi-anode photomultiplier tubes (PMTs) via wavelength-shifting fibers.
The \unit[0.98]{kton} ND is located \unit[1.04]{km} downstream of the
production target and \unit[104]{m} underground. The \unit[5.4]{kton}
FD is approximately \unit[735]{km} downstream of the
target and \unit[705]{m} underground.

Muon neutrinos are identified in the detectors through their
charged-current interactions, $\nu_\mu + A \rightarrow \mu^- + X$ where
$A$ is the target nucleus (normally Fe) and $X$ represents the final
state which may contain pions, other hadrons, and nuclear fragments.  The
muon typically leaves a well-defined energy deposit in the
detector, crossing tens of scintillator planes that can be reconstructed
as a muon track.
Due to the high rate of neutrino interactions within the near detector there are differences
in the way the times of the individual pulses are recorded
in the two detectors. The subsequent reconstruction and selection of the
interactions, however, are done in an identical way. The time and
position of the neutrino interaction vertex is calculated using the TDC
(time-to-digital converter) times of the PMT pulses, combined with the
known spatial geometry of the scintillator strips producing the light.

The resolution on the time of the neutrino interaction is \unit[1.5]{ns} within
both detectors. To achieve this, the detectors are calibrated by first
applying a ``time-walk'' time-slew correction for the variation of time with pulse
height and then a strip-by-strip offset obtained by minimizing the
residual offset in an ensemble of muon tracks. 
The FD time resolution capabilities and
calibration\,\cite{ref:AndyBlakeThesis} were established to distinguish
upward atmospheric neutrino events from downward cosmic
rays\,\cite{ref:minosAtmos2007}, verified in the CalDet test-beam experiment\,\cite{ref:testbeam}, and the detailed calibration constants
were obtained with cosmic ray muon events collected over a number of
years. The ND calibration was performed in a similar way using secondary
muons from the beam and was cross-checked with cosmic rays.  The time
resolution of both detectors is sufficient to resolve
the bunch structure shown in Fig.~\ref{fig:TBeam} and was cross-checked
with portable counters\,\cite{ref:SonCaoThesis}.  A study of the
broadening of the spill structure by the detector resolution was made at
the ND by looking at the reconstructed times of neutrino
interactions within the same accelerator spill to resolve the beam time
structure. 

The MINOS experiment has previously reported a measurement of neutrino
speed in Ref.~\cite{Adamson:2007zzb}.  Since then, the experiment has collected
a factor of 8.5 times more data.  Additionally, a comprehensive study of
the components of the original MINOS timing systems was conducted and
several new corrections have been applied to this larger dataset.  Some
of these studies required components of the new timing system, described
below, to be operated in parallel with the original system; or by
specific tests, comparing measurements between signals in the new and
original systems.  In particular, it was found that a
random offset on the order of \unit[20]{ns} was introduced each time the
original GPS receivers were powered on, which is not unknown in GPS
receivers not specifically designed for precise time.  An offset of this size was
covered by the systematic errors quoted in\,\cite{Adamson:2007zzb}.  The
analysis of the full MINOS data sample, using the original MINOS timing
system and new corrections, yields a systematic error dominated
fractional neutrino speed of $(v/c-1)= (0.6 \pm 1.3) \times 10^{-5}$.

The order-of-magnitude more precise measurement
described in this paper uses a new timing apparatus, making it
insensitive to the variations of the old system.  The rest of this paper
describes this new system and a new analysis based only on the data
taken after the new system was installed.  

\subsection{Proton Beam Measurement}

The time profile of the proton beam is measured using a resistive wall
current monitor (RWCM)\,\cite{ref:RWCM} situated along the beam pipe,
which is between the extraction point from the Main
Injector\,\cite{slipstacking} and the NuMI target. The RWCM consists of
a resistive network bridged across an electrically insulating ceramic
break in the stainless-steel pipe.  As the beam passes, an image current
is induced on the surface of the pipe, creating a measurable voltage
across the resistive network. The voltage signal from the device is
measured for each spill with a waveform digitizer with \unit[1.5]{GHz}
analog bandwidth, which is the limiting bandwidth of the system.

\subsection{Event Timing}

\begin{figure*}[tb]
\centering
\begin{tikzpicture}
\usetikzlibrary{snakes}


\tikzstyle{newcomp}=[draw, rounded corners]
\tikzstyle{oldcomp}=[draw, rounded corners]
\tikzstyle{optical}=[double]
\tikzstyle{distribution unit}=[line width=6, gray!30]
\tikzstyle{dot}=[circle,fill=black,minimum size=1.5pt]

\node [ultra thick, draw] (RWCM) at (2.4,2) {{\bf RWCM}};
\draw[ultra thick, >-] (0,2) -- (RWCM);  \node [below, align=center] at (0.8,2) {{\bf proton}\\{\bf beam}};
\draw[ultra thick, ->] (RWCM) -- (3.5,2);
\draw[line width=6, black] (3.5,2) -- (4,2); \node [above] at (3.75,2.255) {Target};
\draw[ultra thick] (3.5,1.75) rectangle (6,2.25); \node [below] at (4.75,1.75) {Decay tunnel};
\draw[ultra thick, ->] (6,2) -- (8.5,2); \node [below, align=center] at (7,2) {{\bf $\nu_\mu$ beam}\\{\bf 1\,km}};
\draw[ultra thick] (8.5,1.7) rectangle (10.5,2.3); \node at (9.5,2) {{\bf ND}};
\draw[ultra thick, ->] (10.5,2) -- (14,2); \node [below,align=center] at (12,2) {{\bf $\nu_\mu$ beam}\\{\bf 734\,km}};
\draw[ultra thick] (14,1.7) rectangle (17,2.3); \node at (15.5,2) {{\bf FD}};

\draw (1,10.2) -- node[above] {RWCM} (7,10.2);
\draw (7.5,10.2) -- node[above] {Near detector} (12.5,10.2);
\draw (13,10.2) -- node[above] {Far detector} (18,10.2);

\draw[distribution unit] (7.8,3.5) -- (10.5,3.5);
\draw[distribution unit] (13.3,3.5) -- (16,3.5);
\draw[distribution unit] (7.8,8) -- (10.5,8);
\draw[distribution unit] (13.3,8) -- (16,8);
\draw[distribution unit] (2.8,3.5) --(5.5,3.5);

\node[newcomp] (NDTWTTAg) at (10,4) {T};
\draw[optical,->] (10.3,3.5) -- 
   node[right, align=left,pos=0.7] {Underground\\to surface\\TWTT at
     ND} 
   node[right, align=left, pos=0.3] {104\/m\\below\\ground} 
  (10.3,8);
\draw[optical,->] (10,8) -- (NDTWTTAg);
\draw[->] (10,3.5) -- (NDTWTTAg); 

\node[newcomp] (FDTWTTAg) at (15.5,4) {T};
\draw[optical,->] (15.8,3.5) -- 
   node[right, align=left, pos=0.7] {Underground\\to surface\\TWTT at
     FD} 
   node[right, align=left, pos=0.3] {705\/m\\below\\ground} 
   (15.8,8);
\draw[optical,->] (15.5,8) -- (FDTWTTAg);
\draw[->] (15.5,3.5) -- (FDTWTTAg); 

\node[newcomp] (RWCMTWTTAg1) at (8.4,7.5) {T};
\draw[optical,->] (5.4,3.5) -- (5.4,6.8) -- node[below,align=center]
{TWTT between\\RWCM and ND} (8.4,6.8) -- (RWCMTWTTAg1);
\draw[->] (8.4,8) -- (RWCMTWTTAg1); 
\node[newcomp] (RWCMTWTTAg2) at (5.1,4) {T};
\draw[optical,->] (8.1,8) -- (8.1,7.1) -- (5.1,7.1) -- (RWCMTWTTAg2);
\draw[->] (5.1,3.5) -- (RWCMTWTTAg2); 

\node[newcomp, align=center] (NDCs) at (11.5,3.5) {Cs\\clock};
\node[newcomp, align=center] (FDCs) at (17,3.5) {Cs\\clock}; 
\node[newcomp, align=center] (MI60Rb) at (6.8,3.5) {Rb\\clock};
\draw [->] (NDCs) -- (10.5,3.5);
\draw [->] (FDCs) -- (16,3.5);
\draw [->] (MI60Rb) -- (5.5,3.5);

\node[newcomp, align=center] (wavedig) at (1.7,4.0) {Waveform\\Digitizer};
\fill (0.5,3.2) circle(2pt);
\draw[->] (0.5,3.2) |- (wavedig);
\draw[->] (RWCM) |- (1.8,3.4) -| (wavedig);
\draw[->] (3.0,3.5) -- (3.0,4.0) -- (wavedig);
\draw[>-] (0,3.2) -- (0.5,3.2);
\node[below right, align=left] at (0,3.2) {accelerator\\trigger signal};

\foreach \gps/\x/\y in {GPSN1/8.6/8.8, GPSN2/9.7/8.8, GPSF1/14.1/8.8, GPSF2/15.2/8.8, GPSM1/3.4/8.8, GPSM2/4.5/8.8}
 { \node[newcomp] (\gps) at (\x,\y) {GPS};
   \draw (\gps) -- +(0,1.1) -- +(0,0.6) -- +(-0.3,0.9) -- +(0,0.6) -- +(+0.3,0.9);
 }
\draw (GPSN1) -- (8.6,8);
\draw (GPSN2) -- (9.7,8);
\draw (GPSF1) -- (14.1,8);
\draw (GPSF2) -- (15.2,8);
\draw (GPSM1) -- (3.4,3.5);
\draw (GPSM2) -- (4.5,3.5);

\node [oldcomp, align=center] (oldND) at (9,0.5) {Original\\ND timing}; \node at (9,-0.15) {53.1\,MHz};
\node [oldcomp, align=center] (oldFD) at (15.7,0.5) {Original\\FD timing}; \node at (15.7,-0.15) {PPS, \unit[40]{MHz}};
\node [oldcomp, align=center] (oldGPS) at (14,0.5) {Original\\GPS};
\draw [->] (oldGPS) -- (oldFD);

\node[newcomp] (FDdetAg) at (13.7,2.8) {T};
\draw[->] (oldFD) |- (13.7,1.3) -| (FDdetAg);
\draw[->] (13.7,3.5) -- (FDdetAg);
\node[newcomp] (FDdet2Ag) at (15.2,2.8) {T};
\draw[->] (15.2,3.5) -- (FDdet2Ag);
\draw[->] (15.2,2.3) -- (FDdet2Ag); 
\draw[->] (oldFD) -| (16.8,1.7);

\draw[->] (0.5,3.2) -| (6.2,2) |- (oldND);
\node[newcomp] (NDdetAg) at (8.2,2.8) {T};
\draw[->] (8.2,3.5) -- (NDdetAg);
\draw[->] (oldND) |- (8.3,1.3) -| (NDdetAg);
\draw[->] (oldND) -| (10.2,1.7);

\node[newcomp] (RWCMattach) at (3.95,8.2) {X};
\draw (3.95,3.5) -- (RWCMattach);
\node[newcomp] (NDattach) at (7.9,8.4) {X};
\draw (7.9,8) -- (NDattach);
\node[newcomp] (FDattach) at (13.4,8.4) {X};
\draw (13.4,8) -- (FDattach);

\node[right] at (0,0.7) {Key:};
\node[newcomp] (Agilent) at (0.5,0.3) {T};  \node[right] at (1.1,0.3) {Interval timing unit};
\draw[->] (0,0.3) -- (Agilent);
\draw[->] (1,0.3) -- (Agilent);
\draw[distribution unit] (0,-0.1) -- (1,-0.1); \node[right] at (1.1,-0.1) {Distribution unit};
;5A;5A;5A;5A\draw[optical] (0,-0.5) -- (1,-0.5);               \node[right] at (1.1,-0.5) {Optical link};
\node[newcomp] (External) at (0.5,-0.9) {X};     \node[right] at (1.1,-0.9) {Connection point};
\draw (0,-0.9) -- (External);

\end{tikzpicture}
\caption{Simplified layout of the main time synchronization components
  of the experiment. The beam and detectors are shown across the middle
  of the diagram, the newly added time measurement apparatus is shown
  above the beam and original time synchronization (which is still used)
  is shown underneath the beam.  Points marked ``T'' show where an
  interval timer is used permanently and points marked ``X'' show a
  connection point where equipment is connected for short periods during
  the run (the traveling GPS units and the TWSTT (Two-Way Satellite Time
  Transfer) equipment).  Many of
  the links are synchronized/monitored with both PPS and \unit[10]{MHz}
  signals using TWTT (Two-Way Time Transfer).  Not shown are the portable detectors.
  \label{fig:Layout} }
\end{figure*}
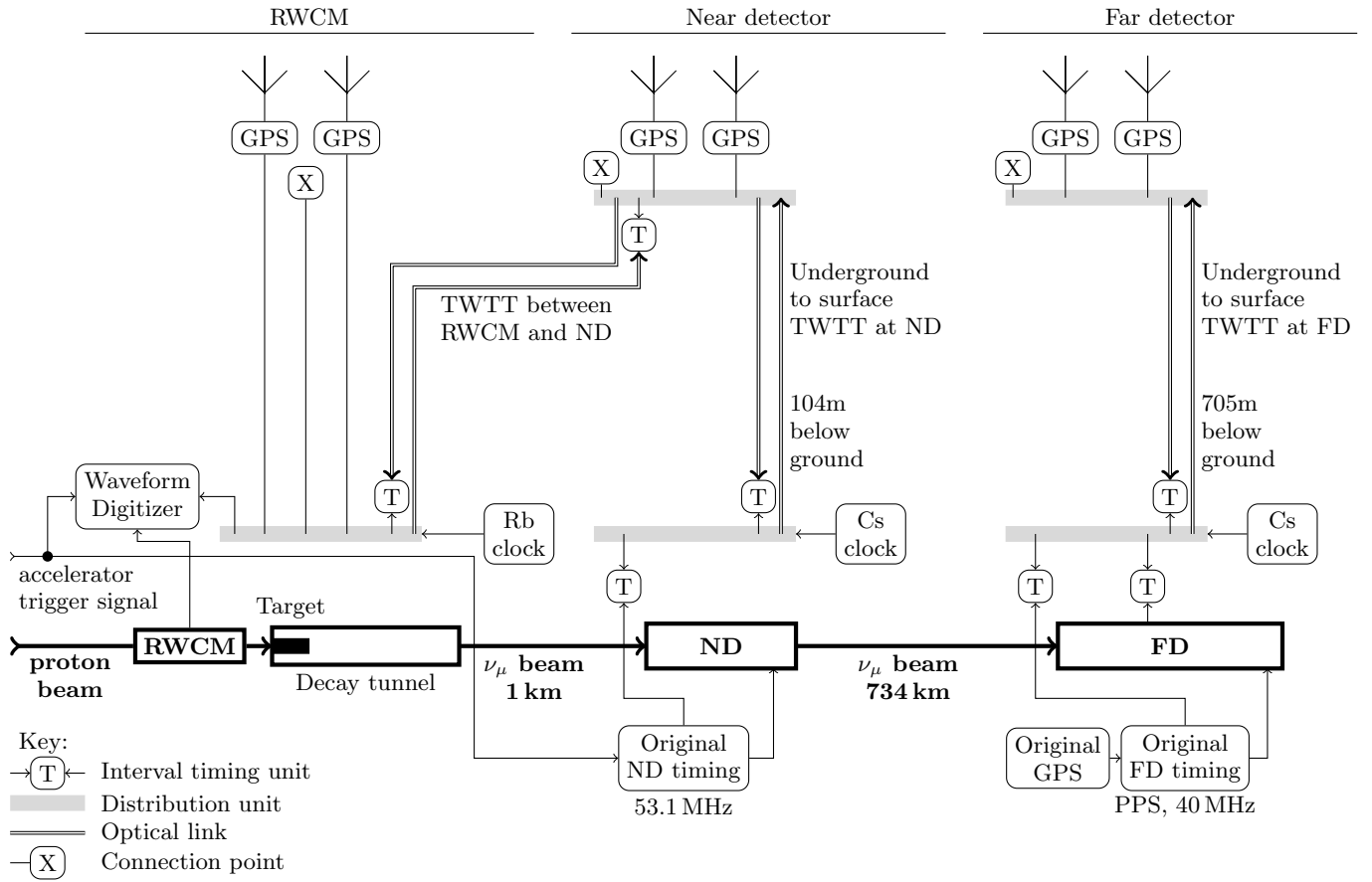

The data presented here use timing components shown in
Fig.~\ref{fig:Layout}.  Original local timing systems used to
internally synchronize different parts of each detector are
retained, as they are integral to the experiment's data acquisition
system.  The new timing system is used to time stamp the old system's
timing markers, allowing more precise calculations of neutrino interaction
times without disturbing the well-tested and robust means of acquiring
the data.  The new system's GPS units and its overall synchronization
are significantly upgraded, as shown in the upper part of
Fig.~\ref{fig:Layout}.  Stable atomic reference clocks are installed at
each detector location.  The manner in which timing synchronization is
transferred between the surface and the underground detector locations
is also upgraded with optical fibers operating in both directions,
transferring both \unit[1]{Hz} (or pulse-per-second, PPS) and
\unit[10]{MHz} signals and allowing continuous monitoring of the delays
in the links.

A local Cs atomic clock is installed at each detector, and a Rb
clock is installed at the RWCM site to establish free running PPS and
\unit[10]{MHz} time references at each site.  Distribution amplifiers
(high output-to-output isolation and low-jitter signal fan-outs) are used to define the reference points and
provide multiple ports to make timing measurements. The internal time
synchronization of each detector is measured with respect to these
timing reference points using interval timers.  A number of these
interval timers are installed permanently in the experiment and used for
the duration of the time-of-flight experiment as shown in
Fig.~\ref{fig:Layout}.  The offsets between the three atomic clocks at
the different sites are continuously measured with the new system of GPS
receivers and interval timers.

At the FD, the time synchronization within the detector uses a
\unit[40]{MHz} clock and PPS boundary markers derived from the original
FD GPS receiver, which runs independently of the new timing equipment.
The PPS are encoded on the \unit[40]{MHz} clock signals to allow
distribution to the 46~readout boards in the detector over a single
network of same-length cables.  The times of the scintillator pulses in
the detectors are measured using a TDC implemented by multiplying the
\unit[40]{MHz} to give \unit[160]{MHz}, then using four delayed
versions of this clock to generate an effective \unit[640]{MHz} clock.

The offset between the new local Cs clock and the PPS from the original
FD GPS receiver (used for the internal detector time synchronization) is
measured and recorded each second with an interval timer.  For each
neutrino interaction, for which data acquisistion is started using a
simple activity trigger, we correct the measured detector time to the
new local Cs clock time using the interval timer measurement closest in
time to the neutrino event.  Variations in the frequency of the
\unit[40]{MHz} FD clock are monitored by measuring the interval between
successive PPS signals with the Cs clock which is observed to vary
smoothly by typically \unit[$\pm10$]{ns} in the short term.  As a
cross-check of the stability of the timing distribution system, signals
were read from one of the readout boards and compared with the new time
reference system using an interval timer.

At the ND, the acquisition of data is started by a trigger signal from
the accelerator which arrives about \unit[20]{$\mu$s} before the
neutrino beam.  The arrival time of this trigger signal within the ND
time distribution electronics is measured against the local Cs clock
reference point with an interval timer. The ND measures the time of the
PMT pulses from neutrino interactions relative to the trigger signal
using a local \unit[53.1]{MHz} clock from a crystal oscillator that is
distributed to the readout boards at the ND.  Cross-checks were also
made with interval timer measurements between points in the old and new
timing systems.

The \unit[53.1]{MHz} clock was found to have a \unit[100]{Hz} variation
synchronous to the accelerator cycle.  However, this clock is only used to
measure short time intervals of $\mathcal{O}(\unit[20]{\mu\mathrm{s}})$,
so the effect is around \unit[40]{ps} at worst and is thus neglected.

Similarly, at the RWCM, the same \unit[20]{$\mu$s} accelerator signal 
used to start digitization at the ND is also used to trigger the RWCM
digitizer and is measured against the Rb clock with an interval timer.
The early accelerator trigger signals at both the ND and RWCM are
derived from the same source, although this is not necessary for the
time-of-flight measurement.  This is simply a signal which precedes the
arrival of the neutrinos by a small amount, used to trigger the
detectors, and measured with respect to the local timing reference point
by the interval counter.

%
%

\subsection{Detector Time Offset}

The mechanisms by which the detectors operate cause systematic time
delays (latencies) between the neutrino interaction time in the detector
and the time when it is recorded.  A separate portable detector is used
to measure the latency of each MINOS detector. The portable detector
consists of a pair of planes, each of active area $\unit[63]{cm}\times
\unit[57]{cm}$
constructed from eight plastic-scintillator strips left over from the
original MINOS construction plus eight similar \unit[3]{cm} wide newly
constructed strips.  The two parallel planes are stacked and oriented so their
strips are rotated by $90^\circ$ from each other.
Coincident signals from the two planes are timed with respect to the
PPS from the local Cs clock reference point.

The portable detector was first placed immediately behind the ND where
some of the muons created in the ND volume by beam neutrinos could pass
through it. By matching PMT pulses in time and correcting for
longitudinal position, we obtain a relative latency measurement between
the portable detector and the ND.  This is measured to be
\unit[$(36\pm4)$]{ns}.  Following this, the portable detector was
transported to the FD and by using cosmic ray muons which passed through
both the FD and the portable detector, the relative latency between the
portable detector and FD was measured to be \unit[$(12\pm 4)$]{ns}.
Most of that up to \unit[4]{ns} unknown latency comes from uncertainties
on the propagation time of the signals in the portable detector itself:
but the same detector and electronics were used at both ND and FD.
Thus, when comparing the ND and FD latencies quoted above, the unknown
delays in the portable detector itself subtract out, leaving a relative
ND-FD latency of \unit[$(24\pm 1)$]{ns}.  The error assigned covers the
jitter on the latency measurements, small drifts observed over time,
differences in magnetic fields at the PMTs at the two locations, and the
somewhat different energies of the different samples of muons used in
the measurement.  A second identical portable detector was used together
with the first one at the ND for studies to characterize the resolution
and stability of the counters and electronics.

%
%
\subsection{Detector and Baseline Survey}

The straight-line distance between the front faces of the near and far
detectors has been determined to \unit[70]{cm} precision.  This section
describes how this precision is achieved.

Survey control networks have been established on the surface and
underground at the ND and FD sites.  Both detectors have been located
relative to their respective underground survey control networks to
within \unit[0.5]{cm}.  The two surface control networks are connected
via high precision GPS measurements to an overall accuracy of about
\unit[1]{cm}.  The tie between the surface network and the ND
underground control network is straightforward and has been accomplished
utilizing standard optical survey methods with millimeter accuracy.
These measurements are described in
Refs.~\cite{ref:minosnim,Bocean:1999bp,Bocean:2000}.

For the FD however, there is no direct plumb line down the sloped shaft and
issues with atmospheric stratification prohibit optical surveys.
Therefore, a Honeywell Inertial Navigation Unit (INS) containing three
gyroscopes and three accelerometers was utilized to connect the surface
and underground control networks.  The INS was
mounted in the elevator cage and traveled multiple times up and down
the mine shaft, stopping each time at four approximately equal distance
positions to reset accumulated velocity errors in the INS.
Limiting factors of the accuracy include: the relatively high vibration
rate of the elevator cage; the fact that the cage stops at slightly
different places each time; and residual oscillations as the cage came
to a stop.  The INS measurement is detailed in Ref.~\cite{ref:inertial}.

Observations from the gyroscopes and accelerometers of the INS were used
to connect the FD's underground coordinates to the surface and
establish the detector's NAD83\,\cite{craymer2000realization}
coordinates.  NAD83 is the horizontal control datum for North America,
and the GRS80\,\cite{mortiz1980geodetic} reference ellipsoid was used in these
conversions.  The position coordinates of the centers of the front faces
of each detector were both converted to these geocentric coordinates in
order to compute the Euclidean distance between the two detectors.  Two
independent coordinate transforms were done, agreeing with each other to
\unit[0.2]{ns}.  The resulting longitudinal uncertainty from the front
face of the ND to the front face of the FD of \unit[70]{cm} is dominated
by the limited INS repeatability measurements in the mine shaft.

The resulting time-of-flight uncertainty caused by the positional
uncertainties in this distance is dominated by the INS error and totals
\unit[2.3]{ns}: this is the dominant systematic error in the final
neutrino speed calculation.  If we instead take the speed of highly
relativistic neutrinos to be given as $c$, we can turn this measurement
around to make a neutrino-based survey of the location of the FD.  This
interpretation of the time-of-flight data presented in the conclusion of
this paper below (Sec.~\ref{sec:analysis}) suggests that the FD is
\unit[(0.72$\pm$0.03${(stat)}\pm$0.39${(syst)}$)]{m} closer to Fermilab
than the inertial survey indicates.

%
%
\section{Time Synchronization}

The time synchronization between sites was implemented using several
independent techniques (GPS, two-way fiber-based and two-way
satellite-based) and also two different processing methods (code-based
common view and carrier-phase based common-view for GPS data)\,\cite{ref:commonview}.
This redundancy ensured robustness and
allowed for assessment of systematics.  The timing systems used in the
experiment are shown schematically in Fig.~\ref{fig:Layout}.  Two GPS
receivers were deployed at each of the three locations (RWCM, ND, and
FD) and a two-way time transfer (TWTT) system using fiber links was
installed between the RWCM and ND.  A two-way satellite time transfer
(TWSTT) synchronization was also performed over a \unit[36]{hour} period
during the data taking period using a dedicated satellite link between
the ND and FD.  The ND and FD atomic clock time references are
underground and were transferred to secondary references on the surface
using TWTT fiber links. The three fiber links in the experiment are
similar and transfer both PPS and \unit[10]{MHz} signals on independent
sets of fibers. All the timing instruments (distribution amplifiers,
optical link transmitters and receivers, 
interval timers) 
were commercial units
specifically intended for transferring or measuring time accurately.
Cables were high quality RG/316-DS using SMA or BNC connectors, and obtained
specifically for this measurement.

\subsection{Fiber links}

The TWTT technique is used in the three fiber links shown on
Fig.~\ref{fig:Layout} by employing a second fiber in the same bundle, so
the fiber is therefore susceptible to the same environmental changes.
The timing signal is sent out from the reference point at the first
location to the second location where it is used to establish a second
reference point.  It is then sent back on the second fiber to the first
location.  At the first location, the delay between the arriving signal
and the first reference point is continuously recorded using an interval
timer and used to correct the data.

The largest variability in the round-trip time is in the FD
surface-underground link, which shows a thermal day/night effect of
\unit[200]{ps} round-trip and a similar overall variability during the
run.  This is corrected to better than \unit[50]{ps} by continuous
monitoring of the round trip time of the PPS along this link.  The
differences in delays of the communication fibers, transmitters and
receivers are determined by swapping modules.

A portable Cs clock was used to verify the calibration of the surface to
underground links by measuring the offset between portable and reference
clocks, first on the surface, then underground, then on the surface
again and correcting for the relative clock drift. The measurements show
an average discrepancy of \unit[400]{ps} at the ND. At the FD, the internal
delays in the optical receivers were not measured directly, but
corrected using the average of a series of three portable clock
measurements made on different days, and the \unit[$\pm550$]{ps} maximum
deviations
of the three measurements is applied as the systematic uncertainty of
this correction.

\subsection{Global Positioning System}

The GPS timing infrastructure consists of eight similar dual-frequency GPS receivers:
six identical receivers (Novatel OEMV) and two newer versions from the
same manufacturer (Novatel OEM6). All receivers use antennas from
Novatel with Andrew FSJ1-50A antenna cables with
small (\unit[-0.028 to +0.036]{ps/m/$^\circ$C}) temperature coeficients.  The
antenna cables were annealed for temperature stability of
the propagation delay before installation.  The group delays of the
antenna cables were measured at both the L1 (\unit[1575.42]{MHz}) and L2
(\unit[1227.60]{MHz}) GPS signal frequencies prior to installation and
confirmed using time-domain reflectometry.
The measured cable delays are used to calculate the time difference
between the two locations.

Two receivers are located at each of the three sites.  The antennas are
located so as to minimize ``multipath'' errors from reflected GPS
satellite signals arriving at slightly later times.  The two remaining
receivers were transported (with their antenna and antenna cable),
between the three MINOS sites and the National Institute of Standards (NIST) in Boulder, CO, to provide
multiple differential calibrations of the fixed GPS systems.  The two
mobile receivers were used at all the sites in different orders.  They were moved
at approximately weekly intervals, spending about three active days at
each site.  When two co-located GPS receivers are calibrated, the
difference between the local time as measured by each receiver is
corrected for the internal delays of the two receivers.  The GPS
timing infrastructure is described in\,\cite{stefaniaPTTI,philPTTI}.

At each location, the timing reference signal from the atomic clock is
input to the GPS receivers.  Each receiver makes measurements of the
time offsets between the timing reference signal and the time obtained
from the GPS satellites which are visible. The GPS data were processed in
Common-View (CV) mode using GPS L1 C/A code-based data to compute the
calibrated differences between the atomic clocks at the various
locations\,\cite{stefaniaPTTI}.  The processing of carrier
phase-based data was done via PPP (precise point positioning) algorithms
that also provide an accurate position of each GPS antenna through the
corrections computed by the International GNSS Service (IGS).  The GPS
data reductions were done by the authors at the National Institute of
Standards and Technology (NIST) using Natural Resources Canada's
CSRS-PPP online service\,\cite{ref:csrs-ppp} and
confirmed by the authors at the US Naval Observatory (USNO) using both
NRCan and the JPL automated data reduction service, which has
independent software (GIPSY/OASIS\,\cite{webb1995introduction}). These reductions provide a large standard
set of corrections including relativistic effects and the ``IGS Final''
corrections for the orbit and atomic-clock time of the satellite.

\begin{figure}[bt]
\centering
\includegraphics[width=8.5 cm]{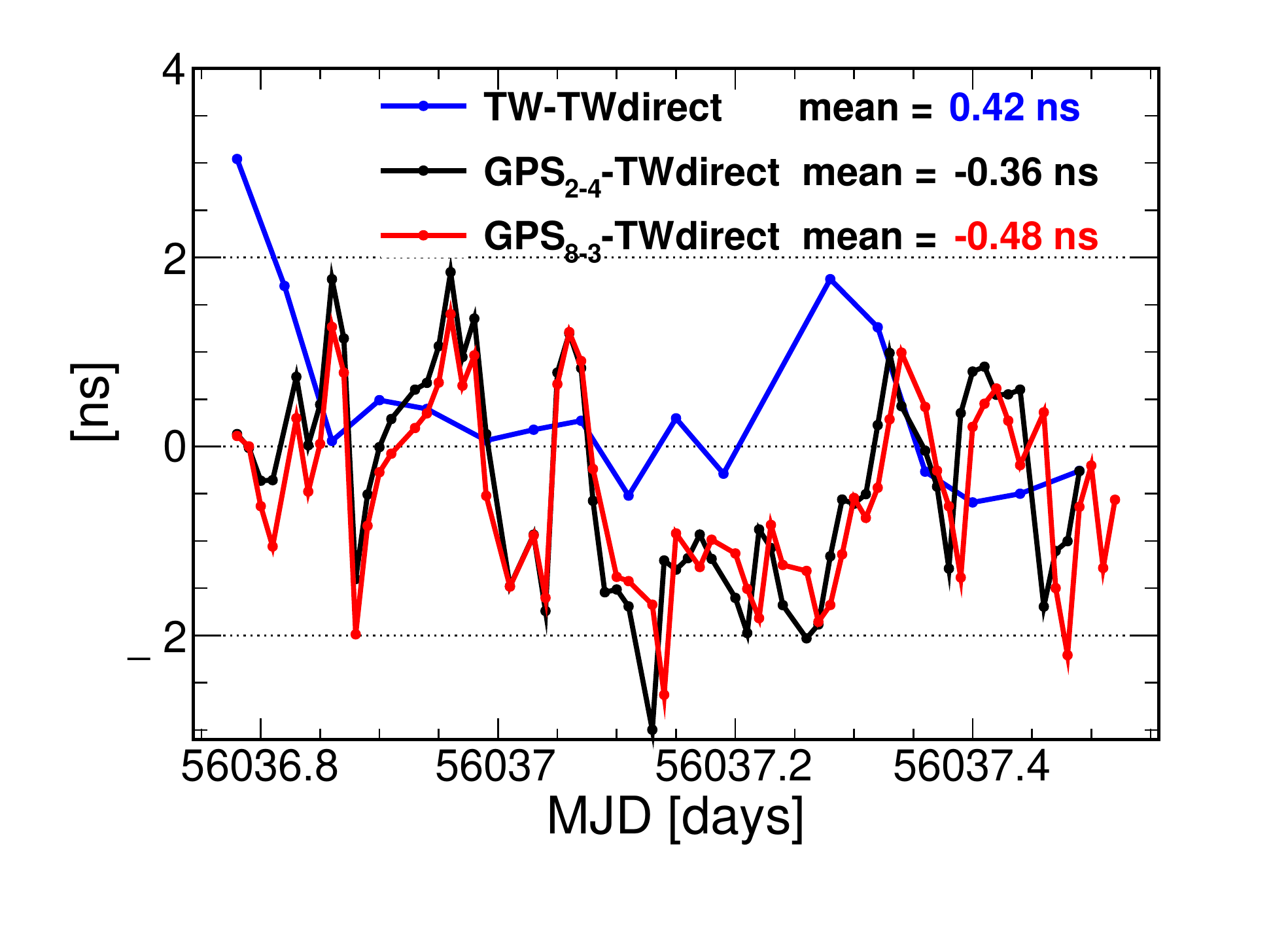}
\caption{ Comparison between TWSTT and the GPS system between ND and FD,
  during the USNO TWSTT test (April 18-19, 2012), expressed as Modified
  Julian Dates (MJD) in the figure.  In the legend, ``TWdirect'' refers
  to calibrated TWSTT between antennas at ND and FD, while ``TW'' refers
  to the double-difference obtained from subtracting observations of
  ND-USNO from FD-USNO.  The subscripts below GPS identify the
  individual GPS receivers used.  The mean differences used in the
  analysis are shown.
\label{fig:TWSTT}
}
\end{figure}

The CV code-based processing of the difference between two co-located
receivers gives an estimate of the hardware stability (GPS receiver,
antenna, and cable).  The time differences over the time-of-flight
measurement period showed better than \unit[200]{ps} (1$\sigma$)
stability at the RWCM and ND.  The GPS receivers and other timing
equipment at the FD are in an environmental chamber
that holds the temperature to \unit[$\pm1$]{K}.  This results in a time
stability of better than \unit[90]{ps}, estimated using the RMS and
the total time deviation\,\cite{stefaniaPTTI}.

\subsection{Synchronization Between Sites}
\label{sec:SiteSynch}

An estimate of the stability and accuracy for the actual
synchronization between sites is obtained by differencing data from
independent time-transfer links between each pair of sites, and where
possible between independent units at the same site.
Effectively this is calculating a double-difference, since each link is
already a difference between the pair of clocks at each location.  Each
technique and each deployed system can have a bias due to a variety of
causes, such as multipath reflected signals, temperature, humidity, and
mismodeling of the atmosphere, satellite orbit, or ionosphere.  These errors can occur
during the initial link calibration, and can also vary over
time.  Double-differences wherein the two links are based on different
techniques (i.e. GPS CV and TWSTT or GPS CV and fiber-based TWTT) are
particularly useful in constraining the biases, since they are
expected to be independent.

A dedicated two-way satellite link was used by US Naval Observatory (USNO) personnel to measure
the time difference between ND and FD references using TWSTT.  This
involves exchanging signals between locations on a bi-directional
satellite link, so that environmental and atmospheric effects are almost
the same in each direction.  The time difference calculated by
double-differencing time-transfer between ND and FD with one GPS-based
link and with the TWSTT link shows stability better than \unit[800]{ps}
and a mean difference (accuracy) of \unit[-480]{ps}.  A secondary TWSTT
mode was also employed using the USNO's facility in Washington DC as a
transfer point, and measuring FD-ND by the double-difference between
USNO-ND and USNO-FD.  No truly relativistic corrections are required in
the TWSTT analysis apart from the formalism to correct for the rotation
of the Earth (the Sagnac correction)\,\cite{Sagnac}.
Figure~\ref{fig:TWSTT} shows the comparison between these time transfer
techniques.

The same double difference computed between one GPS-based link and the
TWSTT link can be performed between two GPS-based links, one using a
pair of GPS receivers (one at each end of the link) and the other using
the other pair. The comparison of the two GPS receivers at each location
also allows the detection of eventual discrepancies between them.
Since the calibration method simultaneously sets all GPS links
between the two sites, the GPS-only double-differences constrain 
the calibration variation between calibrations.  On the ND to FD link,
the stability over the entire neutrino data collection period is better
than \unit[200]{ps} ($1\sigma$). The mean difference of \unit[205]{ps},
suggests a systematic uncertainty of \unit[150]{ps} (half the
error attributed to each link).  We apply an overall systematic error of
\unit[500]{ps} due to the uncertainty in the overall FD-ND
synchronization during the neutrino data collection period.

By using the two traveling GPS units it is possible to perform repeated
calibrations of the stationary receivers at the various locations.  Use
of more than one calibration and the presence of a second receiver at
each location allows the determination of possible time steps in any of the
receiver's internal time base.  As an example, the double-difference
between ND and FD, when calculated using the result of the same
calibration over approximately eight months, showed a step of \unit[2]{ns}.
When the values of subsequent calibrations were considered, it became
clear that the internal time base of one of the stationary receivers at
the ND had a step of about the same magnitude.  Two different
calibration values were thus used for this receiver to remove the step
from the double-difference, eliminating it from the error analysis for
the synchronization.

\section{Data Analysis}
\label{sec:analysis}
%
%

We select two categories of events whose selection criteria are
described in Ref.~\cite{ref:minosCC2010}: contained $\nu_\mu$ charged-current (CC)
events, which originate in the fiducial region of each detector, and,
for the FD only, partially reconstructed events.  Partially
reconstructed events are either $\nu_\mu$ CC neutrino interactions in the rock
surrounding the FD producing an entering muon, or events where the
neutrino interaction point occurred outside the detectors' fiducial region.
Partially reconstructed events usually arrive later than contained
events, as the neutrino produces a muon at an angle, resulting in a
trigonometric increase to the muon path-length (the muons are highly
relativistic and their slightly slower-than-light speed does not contribute significantly to
the later arrival).  Both samples are subject to a further track-fit
quality cut to ensure that we only use events with a well-measured
interaction time. During March and April 2012, an exposure of $0.8\times10^{20}$
protons on target was collected, yielding 195 fully contained and 177
partially reconstructed events which are selected at the FD.

The data analysis proceeds by using the measured time distribution
of the protons in the RWCM to form a
likelihood distribution for the time-of-flight of each neutrino event
to each detector. A likelihood distribution is required because, for a
given neutrino, we do not know which part of the accelerator spill (as
shown in Fig.~\ref{fig:TBeam}) is responsible for
producing the observed neutrino. The RWCM timing waveform is convolved with a
\unit[1.5]{ns} (RMS) wide Gaussian distribution representing the detector
timing resolution and shifted by the predicted time-of-flight to form
a probability distribution function for the contained events. For
partially reconstructed events, the probability density function is further convolved with a
delay distribution to take account of the increased muon path length
calculated using Monte Carlo simulations. 

We multiply the event likelihoods of all the observed events together to
obtain an overall probability as a function of the time-of-flight
parameter. The time-of-flight which gives the maximum combined
probability is then our measurement of the overall neutrino flight time
from the RWCM to the detector.  This procedure is followed at each of
the two detectors.

To obtain the final result for the ND to FD time-of-flight, the two
times obtained are subtracted, eliminating any systematic offset
associated with the beam measurement and the portable detector
measurement.  Having two neutrino detectors, near and far, is unique to
this measurement, as other neutrino time-of-flight analyses must rely on
beam measurements.  

\begin{figure}[tb]
\centering
\includegraphics[width=8.5 cm]{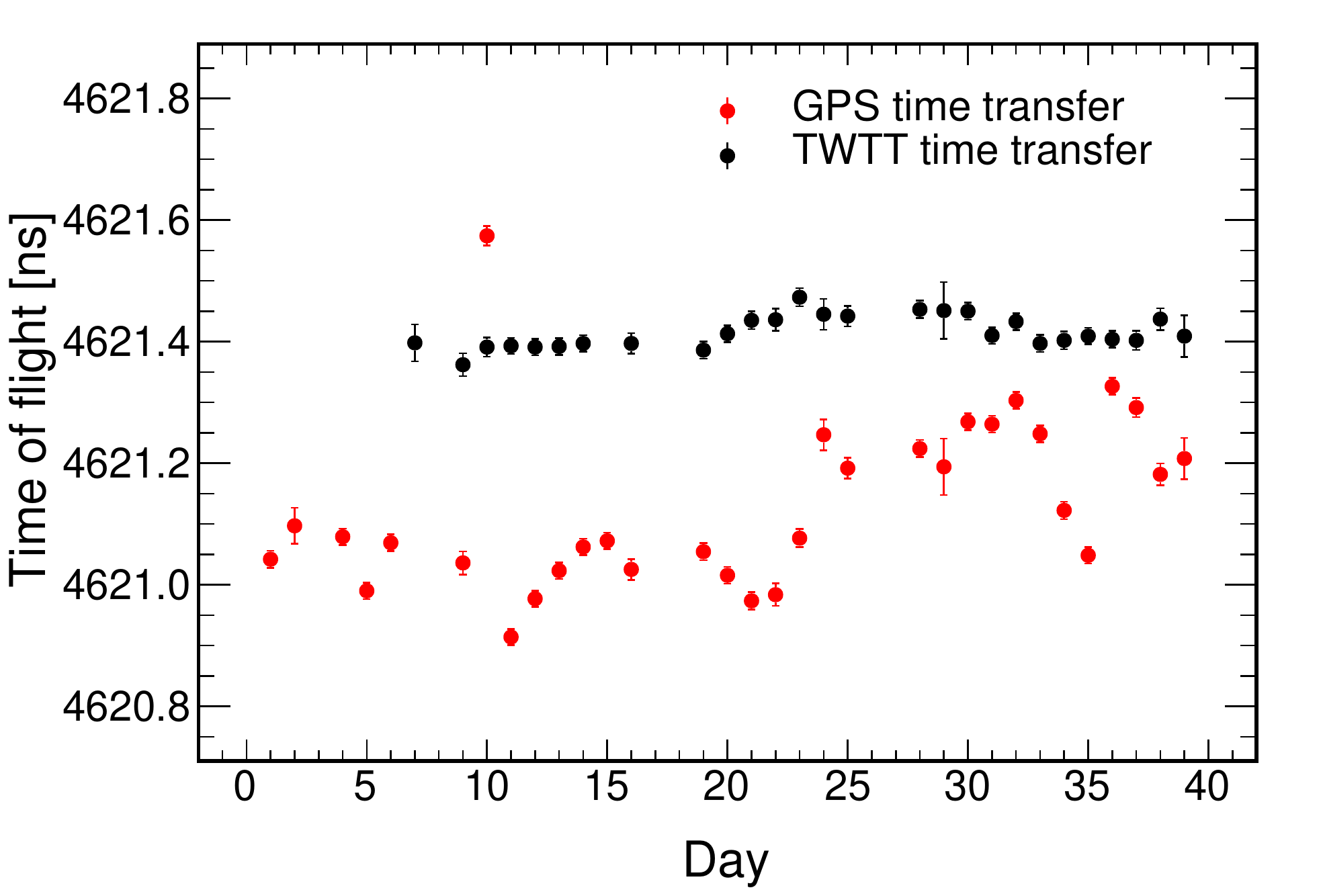}
\caption{
  Daily variation in time-of-flight between the RWCM and ND. The GPS (red
  points) and TWTT (black points) techniques were used for transferring the
  time between the two sites.  The difference between these two
  techniques is confirmation of the stability of the time synchronization.
\label{fig:NDDaily}
}
\end{figure}

At the ND, the neutrino interaction rate is sufficient to measure the
time-of-flight for each day with statistical error below \unit[50]{ps},
permitting a test of the long-term stability of both the GPS and the
TWTT timing systems using neutrinos directly, over the short baseline
between RWCM and ND.  The results from both these methods are shown in
Fig.~\ref{fig:NDDaily}.  The surveyed distance between the RWCM and the
ND is combined with the absolute latency of the ND measured with the
portable detector to create an expected time-of-flight of \unit[$4622.7
\pm 4.0$]{ns}, where the uncertainty comes from the absolute latency of
the portable detector and the RWCM-ND synchronization.  Also included is
a small effect of $\mathcal{O}(\unit[500]{ps})$ from the protons and pions
traveling slightly below the speed of light and away from the beam axis.

\begin{table}[tb]
\caption{Dominant systematic uncertainties ($\pm\mathrm{1}\sigma$) \label{tab:syst} }
\begin{tabular}{l c} 
\hline
Systematic Uncertainty & Value\\
\hline
Inertial survey of the FD location & \unit[2.3]{ns} \\
Relative ND-FD latency & $1.0$\,ns \\
FD TWTT between surface and underground & \unit[0.6]{ns} \\
GPS time-transfer accuracy & \unit[0.5]{ns} \\
\hline
\end{tabular}
\end{table}

Figure~\ref{fig:NDDaily} shows that the daily measurements from both
methods are contained within a \unit[1]{ns} range consistent with this
expectation.  Note that the precision of this test is not as good as on
the final time-of-flight result which benefits from the cancellation of
uncertainty in the subtraction technique.  Figure~\ref{fig:NDDaily}
shows that the time-of-flight measurements are stable to about
\unit[200]{ps} when using the GPS for time-transfer and to better than
\unit[50]{ps} when using the TWTT for time transfer.  As the GPS time
was common to both detectors and crosschecked by the USNO TWSTT
measurement, the mean GPS-based
value of \unit[4621.1]{ns} is used later in calculating the final
time-of-flight, with the difference being covered by the systematic
error described in Sec.\ref{sec:SiteSynch}.

\begin{figure}[tb]
\centering
\includegraphics[width=8.5 cm]{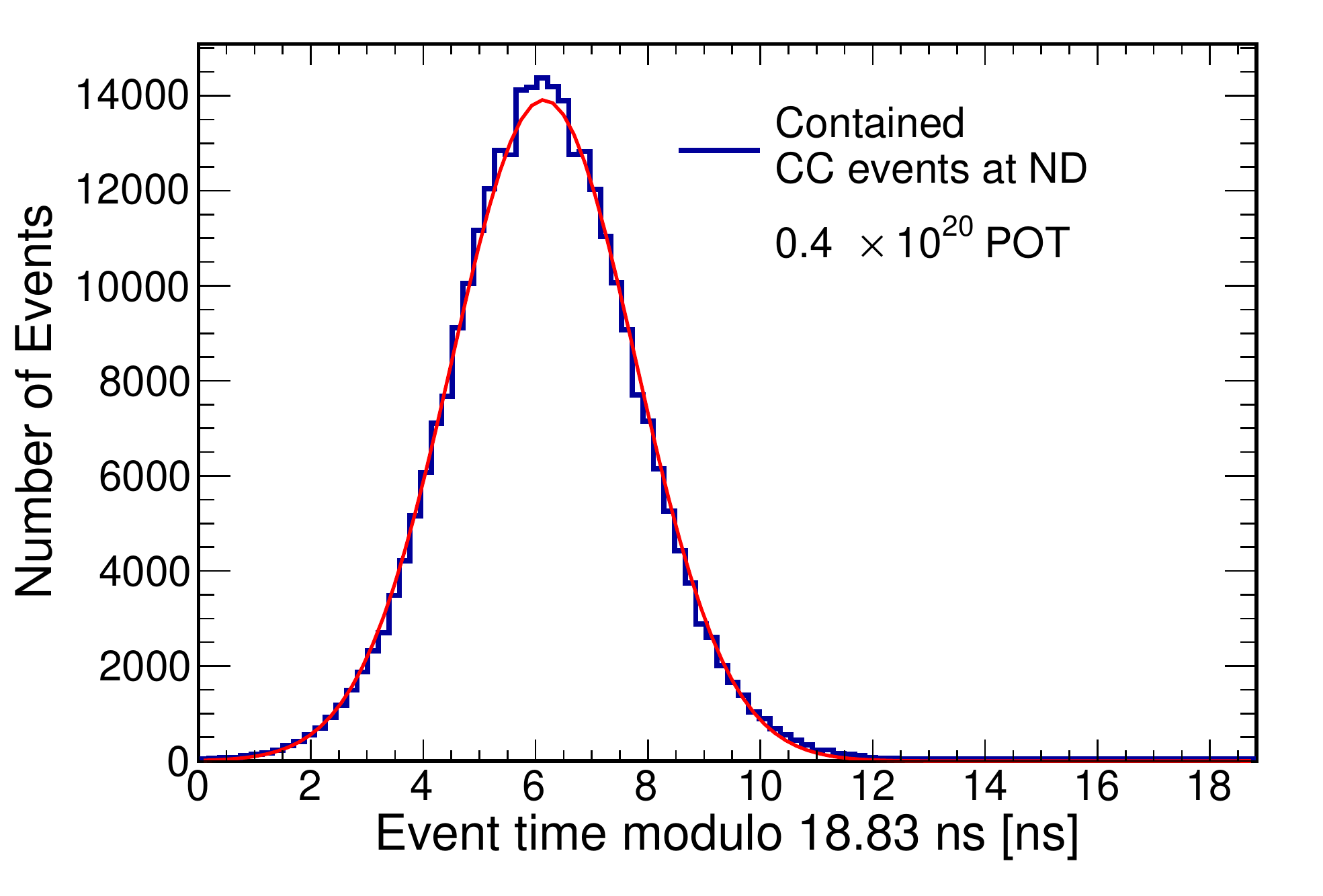}
\caption{Arrival time distribution at the near detector modulo the
  \unit[18.83]{ns} bunch separation (blue histogram), which is well fit
  by a gaussian with a \unit[1.6]{ns} sigma (red line).  The high
  neutrino statistics available at the ND establish the shape and timing
  of the neutrino bunch structure\label{fig:NDbun}.
}
\end{figure}

\begin{figure}[tb]
\centering
\includegraphics[width=8.5 cm]{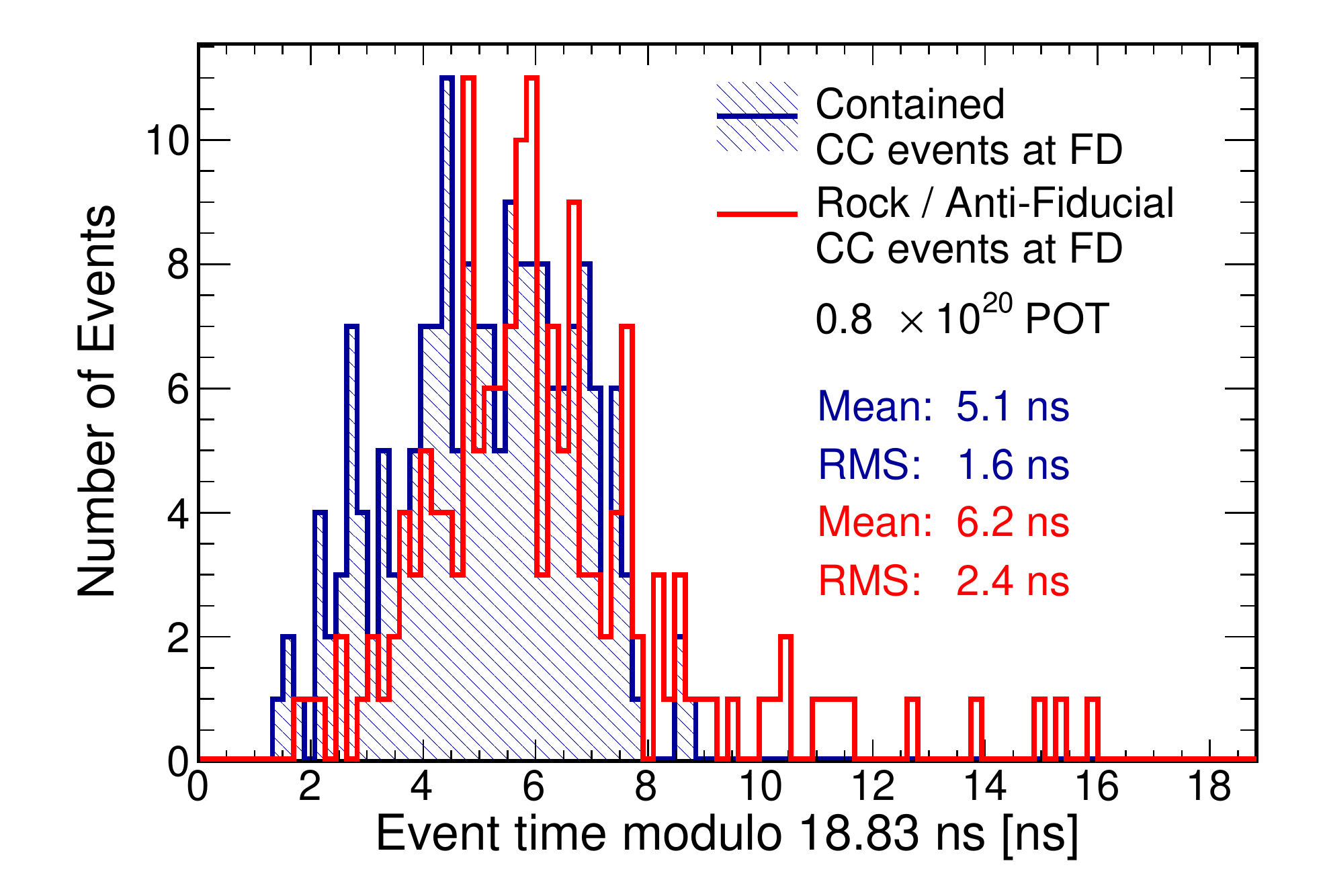}
\caption{Arrival time distribution at the far detector modulo the
  \unit[18.83]{ns} bunch separation for contained (blue hatched histogram) and
  partially contained
  (red line) events. The appearance of the bunch shape in the FD
  demonstrates that the timing system is functioning as
  expected.\label{fig:FDbun}
}
\end{figure}

Figure~\ref{fig:NDbun} shows the neutrino arrival time distribution at
the ND after the end of the preceding \unit[18.83]{ns} long bunch, and
Fig.~\ref{fig:FDbun} the same at the FD.  The high neutrino statistics
available at the ND illustrate that the neutrino production in a bunch
is well fit by a Gaussian with a \unit[1.6]{ns} sigma: this width is
driven by the per-event \unit[1.5]{ns} detector time resolution rather than the
bunch width itself.  The contained
event sample at the FD is consistent with that distribution, confirming
the resolution and stability of the time measurement system.  The
partially reconstructed events at the FD are also shown in
Fig.~\ref{fig:FDbun} and are seen to arrive later and with a bigger
spread as expected.

Combining the contained and partially reconstructed samples, the
time-of-flight between the RWCM and FD is found to be
\unit[$(2\,453\,935.0\pm 0.1)$]{ns}, considering only statistical errors.
Subtracting the measured time-of-flight (using GPS) between RWCM and ND
of \unit[4621.1]{ns} we obtain the time-of-flight between ND and FD as
\unit[$(2\,449\,313.9\pm 0.1)$]{ns} (statistical error only): the most
precise measurement of the neutrino time-of-flight ever achieved, and
the only one obtained directly using two neutrino detectors. The time
required to traverse the distance between the front face of the Near and
Far detectors at the speed of light, including the Sagnac correction, is
\unit[$(2\,449\,316.3\pm 2.3)$]{ns}, where the dominant uncertainty comes
from the inertial survey of the FD location.  Combining these, together
with the other sources of systematic error listed in
Table~\ref{tab:syst}, yields a value for the difference in arrival time
of the neutrino and the speed of light prediction of $\delta = \unit[(2.4
\pm 0.1 (stat.) \pm 2.6 (syst.))$]{ns}.  The
fractional neutrino
speed is therefore found to be $(v/c-1) = (1.0 \pm 1.1) \times
10^{-6}$, consistent to \unit[1]{$\sigma$} with relativistic neutrinos.

\section*{Acknowledgments}

This work was supported by the U.S. DOE; the United Kingdom STFC; the
U.S. NSF; the State and University of Minnesota; Brazil's FAPESP, CNPq
and CAPES.  We are grateful to the Minnesota Department of Natural
Resources and the personnel of the Soudan Laboratory, Fermilab, and USNO.  We thank Texas Advanced Computing Center at The University of
Texas at Austin for the provision of computing resources.  Fermilab is
Operated by Fermi Research Alliance, LLC under Contract
No.~De-AC02-07CH11359 with the United States Department of Energy.

\bibliography{tof}

\end{document}

%% file: TOF-Nov14-authors.tex
%



\newcommand{\Berkeley}{Lawrence Berkeley National Laboratory, Berkeley, California, 94720 USA}
\newcommand{\Cambridge}{Cavendish Laboratory, University of Cambridge, Madingley Road, Cambridge CB3 0HE, United Kingdom}
\newcommand{\Cincinnati}{Department of Physics, University of Cincinnati, Cincinnati, Ohio 45221, USA}
\newcommand{\FNAL}{Fermi National Accelerator Laboratory, Batavia, Illinois 60510, USA}
\newcommand{\RAL}{Rutherford Appleton Laboratory, Science and
  Technologies Facilities Council, Didcot, OX11 0QX, United Kingdom}
\newcommand{\UCL}{Department of Physics and Astronomy, University College London, Gower Street, London WC1E 6BT, United Kingdom}
\newcommand{\Caltech}{Lauritsen Laboratory, California Institute of Technology, Pasadena, California 91125, USA}
\newcommand{\Alabama}{Department of Physics and Astronomy, University of Alabama, Tuscaloosa, Alabama 35487, USA}
\newcommand{\ANL}{Argonne National Laboratory, Argonne, Illinois 60439, USA}
\newcommand{\Athens}{Department of Physics, University of Athens, GR-15771 Athens, Greece}
\newcommand{\NTUAthens}{Department of Physics, National Tech. University of Athens, GR-15780 Athens, Greece}
\newcommand{\Benedictine}{Physics Department, Benedictine University, Lisle, Illinois 60532, USA}
\newcommand{\BNL}{Brookhaven National Laboratory, Upton, New York 11973, USA}
\newcommand{\CdF}{APC -- Universit\'{e} Paris 7 Denis Diderot, 10, rue Alice Domon et L\'{e}onie Duquet, F-75205 Paris Cedex 13, France}
\newcommand{\Cleveland}{Cleveland Clinic, Cleveland, Ohio 44195, USA}
\newcommand{\Delhi}{Department of Physics \& Astrophysics, University of Delhi, Delhi 110007, India}
\newcommand{\GEHealth}{GE Healthcare, Florence South Carolina 29501, USA}
\newcommand{\Harvard}{Department of Physics, Harvard University, Cambridge, Massachusetts 02138, USA}
\newcommand{\HolyCross}{Holy Cross College, Notre Dame, Indiana 46556, USA}
\newcommand{\Houston}{Department of Physics, University of Houston, Houston, Texas 77204, USA}
\newcommand{\IIT}{Department of Physics, Illinois Institute of Technology, Chicago, Illinois 60616, USA}
\newcommand{\Iowa}{Department of Physics and Astronomy, Iowa State University, Ames, Iowa 50011 USA}
\newcommand{\Indiana}{Indiana University, Bloomington, Indiana 47405, USA}
\newcommand{\ITEP}{High Energy Experimental Physics Department, ITEP, B. Cheremushkinskaya, 25, 117218 Moscow, Russia}
\newcommand{\JMU}{Physics Department, James Madison University, Harrisonburg, Virginia 22807, USA}
\newcommand{\LASL}{Nuclear Nonproliferation Division, Threat Reduction Directorate, Los Alamos National Laboratory, Los Alamos, New Mexico 87545, USA}
\newcommand{\Lebedev}{Nuclear Physics Department, Lebedev Physical Institute, Leninsky Prospect 53, 119991 Moscow, Russia}
\newcommand{\LLL}{Lawrence Livermore National Laboratory, Livermore, California 94550, USA}
\newcommand{\LosAlamos}{Los Alamos National Laboratory, Los Alamos, New Mexico 87545, USA}
\newcommand{\Manchester}{School of Physics and Astronomy, University of Manchester, Oxford Road, Manchester M13 9PL, United Kingdom}
\newcommand{\MIT}{Lincoln Laboratory, Massachusetts Institute of Technology, Lexington, Massachusetts 02420, USA}
\newcommand{\Minnesota}{University of Minnesota, Minneapolis, Minnesota 55455, USA}
\newcommand{\Crookston}{Math, Science and Technology Department, University of Minnesota -- Crookston, Crookston, Minnesota 56716, USA}
\newcommand{\Duluth}{Department of Physics \& Astronomy, University of Minnesota Duluth, Duluth, Minnesota 55812, USA}

\newcommand{\NDSU}{North Dakota State University, Fargo, North Dakota
  58108, USA}
\newcommand{\NIST}{NIST, Time and Frequency Division, Boulder, Colorado 80305, USA}  

\newcommand{\Ohio}{Center for Cosmology and Astro Particle Physics, Ohio State University, Columbus, Ohio 43210 USA}
\newcommand{\Otterbein}{Otterbein College, Westerville, Ohio 43081, USA}
\newcommand{\Oxford}{Subdepartment of Particle Physics, University of Oxford, Oxford OX1 3RH, United Kingdom}
\newcommand{\PennState}{Department of Physics, Pennsylvania State University, State College, Pennsylvania 16802, USA}
\newcommand{\PennU}{Department of Physics and Astronomy, University of Pennsylvania, Philadelphia, Pennsylvania 19104, USA}
\newcommand{\Pittsburgh}{Department of Physics and Astronomy, University of Pittsburgh, Pittsburgh, Pennsylvania 15260, USA}
\newcommand{\IHEP}{Institute for High Energy Physics, Protvino, Moscow Region RU-140284, Russia}
\newcommand{\Rochester}{Department of Physics and Astronomy, University of Rochester, New York 14627 USA}
\newcommand{\RoyalH}{Physics Department, Royal Holloway, University of London, Egham, Surrey, TW20 0EX, United Kingdom}
\newcommand{\Carolina}{Department of Physics and Astronomy, University of South Carolina, Columbia, South Carolina 29208, USA}
\newcommand{\SDakota}{South Dakota School of Mines and Technology, Rapid City, South Dakota 57701, USA}
\newcommand{\SLAC}{Stanford Linear Accelerator Center, Stanford, California 94309, USA}
\newcommand{\Stanford}{Department of Physics, Stanford University, Stanford, California 94305, USA}
\newcommand{\StJohnFisher}{Physics Department, St. John Fisher College, Rochester, New York 14618 USA}
\newcommand{\Sussex}{Department of Physics and Astronomy, University of Sussex, Falmer, Brighton BN1 9QH, United Kingdom}
\newcommand{\TexasAM}{Physics Department, Texas A\&M University, College Station, Texas 77843, USA}
\newcommand{\Texas}{Department of Physics, University of Texas at Austin, 1 University Station C1600, Austin, Texas 78712, USA}
\newcommand{\TechX}{Tech-X Corporation, Boulder, Colorado 80303, USA}
\newcommand{\Tufts}{Physics Department, Tufts University, Medford, Massachusetts 02155, USA}
\newcommand{\UNICAMP}{Universidade Estadual de Campinas, IFGW-UNICAMP, CP 6165, 13083-970, Campinas, SP, Brazil}
\newcommand{\UFG}{Instituto de F\'{i}sica, Universidade Federal de Goi\'{a}s, CP 131, 74001-970, Goi\^{a}nia, GO, Brazil}
\newcommand{\USP}{Instituto de F\'{i}sica, Universidade de S\~{a}o Paulo,  CP 66318, 05315-970, S\~{a}o Paulo, SP, Brazil}

\newcommand{\USNO}{USNO, Washington, DC, USA}   

\newcommand{\Warsaw}{Department of Physics, University of Warsaw, Pasteura 5, PL-02-093 Warsaw, Poland}
\newcommand{\Washington}{Physics Department, Western Washington University, Bellingham, Washington 98225, USA}
\newcommand{\WandM}{Department of Physics, College of William \& Mary, Williamsburg, Virginia 23187, USA}
\newcommand{\Wisconsin}{Physics Department, University of Wisconsin, Madison, Wisconsin 53706, USA}
\newcommand{\deceased}{Deceased.}

\affiliation{\ANL}
\affiliation{\BNL}
\affiliation{\Caltech}
\affiliation{\Cambridge}
\affiliation{\UNICAMP}
\affiliation{\Cincinnati}
\affiliation{\FNAL}
\affiliation{\UFG}
\affiliation{\Harvard}
\affiliation{\HolyCross}
\affiliation{\Houston}
\affiliation{\IIT}
\affiliation{\Indiana}
\affiliation{\Iowa}
\affiliation{\UCL}
\affiliation{\Manchester}
\affiliation{\Minnesota}
\affiliation{\Duluth}
\affiliation{\Otterbein}
\affiliation{\Oxford}
\affiliation{\Pittsburgh}
\affiliation{\RAL}
\affiliation{\USP}
\affiliation{\Carolina}
\affiliation{\Stanford}
\affiliation{\Sussex}
\affiliation{\TexasAM}
\affiliation{\Texas}
\affiliation{\Tufts}
\affiliation{\Warsaw}
\affiliation{\WandM}

\author{P.~Adamson}
\affiliation{\FNAL}


\author{I.~Anghel}
\affiliation{\Iowa}
\affiliation{\ANL}



\author{N.~Ashby}\affiliation{\NIST} 

\author{A.~Aurisano}
\affiliation{\Cincinnati}









\author{G.~Barr}
\affiliation{\Oxford}









\author{M.~Bishai}
\affiliation{\BNL}

\author{A.~Blake}
\affiliation{\Cambridge}


\author{G.~J.~Bock}
\affiliation{\FNAL}


\author{D.~Bogert}
\affiliation{\FNAL}




\author{R.~Bumgarner}\affiliation{\USNO} 

% ...

\collaboration{The MINOS Collaboration, NIST, and USNO}
\noaffiliation




%% file: tof.bbl
\begin{thebibliography}{33}
\expandafter\ifx\csname natexlab\endcsname\relax\def\natexlab#1{#1}\fi
\expandafter\ifx\csname bibnamefont\endcsname\relax
  \def\bibnamefont#1{#1}\fi
\expandafter\ifx\csname bibfnamefont\endcsname\relax
  \def\bibfnamefont#1{#1}\fi
\expandafter\ifx\csname citenamefont\endcsname\relax
  \def\citenamefont#1{#1}\fi
\expandafter\ifx\csname url\endcsname\relax
  \def\url#1{\texttt{#1}}\fi
\expandafter\ifx\csname urlprefix\endcsname\relax\def\urlprefix{URL }\fi
\providecommand{\bibinfo}[2]{#2}
\providecommand{\eprint}[2][]{\url{#2}}

\bibitem[{\citenamefont{Hirata et~al.}(1987)}]{Hirata:1987hu}
\bibinfo{author}{\bibfnamefont{K.}~\bibnamefont{Hirata}} \bibnamefont{et~al.}
  (\bibinfo{collaboration}{KAMIOKANDE-II}), \bibinfo{journal}{Phys.\ Rev.\
  Lett.} \textbf{\bibinfo{volume}{58}}, \bibinfo{pages}{1490}
  (\bibinfo{year}{1987}).

\bibitem[{\citenamefont{Bionta et~al.}(1987)}]{Bionta:1987qt}
\bibinfo{author}{\bibfnamefont{R.}~\bibnamefont{Bionta}} \bibnamefont{et~al.}
  (\bibinfo{collaboration}{IMB}), \bibinfo{journal}{Phys.\ Rev.\ Lett.}
  \textbf{\bibinfo{volume}{58}}, \bibinfo{pages}{1494} (\bibinfo{year}{1987}).

\bibitem[{\citenamefont{Alekseev et~al.}(1987)}]{Alekseev:1987ej}
\bibinfo{author}{\bibfnamefont{E.}~\bibnamefont{Alekseev}} \bibnamefont{et~al.}
  (\bibinfo{collaboration}{Baksan}), \bibinfo{journal}{JETP Lett.}
  \textbf{\bibinfo{volume}{45}}, \bibinfo{pages}{589} (\bibinfo{year}{1987}).

\bibitem[{\citenamefont{Kalbfleisch et~al.}(1979)\citenamefont{Kalbfleisch,
  Baggett, Fowler, and Alspector}}]{Kalbfleisch:1979rm}
\bibinfo{author}{\bibfnamefont{G.R.}~\bibnamefont{Kalbfleisch}},
  \bibinfo{author}{\bibfnamefont{N.}~\bibnamefont{Baggett}},
  \bibinfo{author}{\bibfnamefont{E.C.}~\bibnamefont{Fowler}}, \bibnamefont{and}
  \bibinfo{author}{\bibfnamefont{J.}~\bibnamefont{Alspector}},
  \bibinfo{journal}{Phys.\ Rev.\ Lett.} \textbf{\bibinfo{volume}{43}},
  \bibinfo{pages}{1361} (\bibinfo{year}{1979}).

\bibitem[{\citenamefont{Adamson et~al.}(2007{\natexlab{a}})}]{Adamson:2007zzb}
\bibinfo{author}{\bibfnamefont{P.}~\bibnamefont{Adamson}} \bibnamefont{et~al.}
  (\bibinfo{collaboration}{MINOS}), \bibinfo{journal}{Phys.\ Rev.}
  \textbf{\bibinfo{volume}{D76}}, \bibinfo{pages}{072005}
  (\bibinfo{year}{2007}{\natexlab{a}}).

\bibitem[{\citenamefont{Adam et~al.}(2011)}]{Adam:2011}
\bibinfo{author}{\bibfnamefont{T.}~\bibnamefont{Adam}} \bibnamefont{et~al.}
  (\bibinfo{collaboration}{OPERA}) (\bibinfo{year}{2011}),
  \eprint{1109.4897v2}.

\bibitem[{\citenamefont{Adam et~al.}(2012)}]{Adam:2012}
\bibinfo{author}{\bibfnamefont{T.}~\bibnamefont{Adam}} \bibnamefont{et~al.}
  (\bibinfo{collaboration}{OPERA}), \bibinfo{journal}{JHEP}
  \textbf{\bibinfo{volume}{1210}}, \bibinfo{pages}{093} (\bibinfo{year}{2012}),
  \eprint{1109.4897v4}.

\bibitem[{\citenamefont{Antonello
  et~al.}(2012{\natexlab{a}})}]{Antonello:2012hg}
\bibinfo{author}{\bibfnamefont{M.}~\bibnamefont{Antonello}}
  \bibnamefont{et~al.} (\bibinfo{collaboration}{ICARUS}),
  \bibinfo{journal}{Phys.\ Lett.} \textbf{\bibinfo{volume}{B713}},
  \bibinfo{pages}{17} (\bibinfo{year}{2012}{\natexlab{a}}).

\bibitem[{\citenamefont{Agafonova et~al.}(2012)}]{Agafonova:2012rh}
\bibinfo{author}{\bibfnamefont{N.~Y.} \bibnamefont{Agafonova}}
  \bibnamefont{et~al.} (\bibinfo{collaboration}{LVD}), \bibinfo{journal}{Phys.\
  Rev.\ Lett.} \textbf{\bibinfo{volume}{109}}, \bibinfo{pages}{070801}
  (\bibinfo{year}{2012}).

\bibitem[{\citenamefont{Antonello
  et~al.}(2012{\natexlab{b}})}]{Antonello:2012be}
\bibinfo{author}{\bibfnamefont{M.}~\bibnamefont{Antonello}}
  \bibnamefont{et~al.} (\bibinfo{collaboration}{ICARUS}),
  \bibinfo{journal}{JHEP} \textbf{\bibinfo{volume}{1211}}, \bibinfo{pages}{049}
  (\bibinfo{year}{2012}{\natexlab{b}}).

\bibitem[{\citenamefont{Adam et~al.}(2013)}]{Adam:2012pk}
\bibinfo{author}{\bibfnamefont{T.}~\bibnamefont{Adam}} \bibnamefont{et~al.}
  (\bibinfo{collaboration}{OPERA}), \bibinfo{journal}{JHEP}
  \textbf{\bibinfo{volume}{1301}}, \bibinfo{pages}{153} (\bibinfo{year}{2013}).

\bibitem[{\citenamefont{Alvarez~Sanchez et~al.}(2012)}]{AlvarezSanchez:2012wg}
\bibinfo{author}{\bibfnamefont{P.}~\bibnamefont{Alvarez~Sanchez}}
  \bibnamefont{et~al.} (\bibinfo{collaboration}{Borexino}),
  \bibinfo{journal}{Phys. Lett.} \textbf{\bibinfo{volume}{B716}},
  \bibinfo{pages}{401} (\bibinfo{year}{2012}), \eprint{1207.6860}.

\bibitem[{\citenamefont{Adamson et~al.}(2015)}]{Adamson:2015dkw}
\bibinfo{author}{\bibfnamefont{P.}~\bibnamefont{Adamson}} \bibnamefont{et~al.}
  (\bibinfo{year}{2015}), \eprint{1507.06690}.

\bibitem[{\citenamefont{Michael et~al.}(2008)}]{ref:minosnim}
\bibinfo{author}{\bibfnamefont{D.~G.} \bibnamefont{Michael}}
  \bibnamefont{et~al.} (\bibinfo{collaboration}{MINOS}),
  \bibinfo{journal}{Nucl. Instrum. Meth. A} \textbf{\bibinfo{volume}{596}},
  \bibinfo{pages}{190} (\bibinfo{year}{2008}).

\bibitem[{\citenamefont{Adamson et~al.}(2013)}]{ref:combinedNuMu}
\bibinfo{author}{\bibfnamefont{P.}~\bibnamefont{Adamson}} \bibnamefont{et~al.}
  (\bibinfo{collaboration}{MINOS Collaboration}), \bibinfo{journal}{Phys.\
  Rev.\ Lett.} \textbf{\bibinfo{volume}{110}}, \bibinfo{pages}{251801}
  (\bibinfo{year}{2013}).

\bibitem[{\citenamefont{Blake}(2005)}]{ref:AndyBlakeThesis}
\bibinfo{author}{\bibfnamefont{A.}~\bibnamefont{Blake}},
  \bibinfo{journal}{Ph.D. Thesis, U. of Cambridge}  (\bibinfo{year}{2005}).

\bibitem[{\citenamefont{Adamson
  et~al.}(2007{\natexlab{b}})}]{ref:minosAtmos2007}
\bibinfo{author}{\bibfnamefont{P.}~\bibnamefont{Adamson}} \bibnamefont{et~al.}
  (\bibinfo{collaboration}{MINOS}), \bibinfo{journal}{Phys.\ Rev.\ D}
  \textbf{\bibinfo{volume}{75}}, \bibinfo{pages}{092003}
  (\bibinfo{year}{2007}{\natexlab{b}}).

\bibitem[{\citenamefont{Cabrera et~al.}(2009)}]{ref:testbeam}
\bibinfo{author}{\bibfnamefont{A.}~\bibnamefont{Cabrera}} \bibnamefont{et~al.}
  (\bibinfo{collaboration}{MINOS}), \bibinfo{journal}{Nucl.\ Instrum.\ Meth.}
  \textbf{\bibinfo{volume}{A609}}, \bibinfo{pages}{106} (\bibinfo{year}{2009}).

\bibitem[{\citenamefont{Cao}(2014)}]{ref:SonCaoThesis}
\bibinfo{author}{\bibfnamefont{S.}~\bibnamefont{Cao}}, \bibinfo{journal}{Ph.D.
  Thesis, U. of Texas at Austin}  (\bibinfo{year}{2014}).

\bibitem[{\citenamefont{Crisp and Fellenz}(2011)}]{ref:RWCM}
\bibinfo{author}{\bibfnamefont{J.}~\bibnamefont{Crisp}} \bibnamefont{and}
  \bibinfo{author}{\bibfnamefont{B.}~\bibnamefont{Fellenz}},
  \bibinfo{journal}{JINST} \textbf{\bibinfo{volume}{6}},
  \bibinfo{pages}{T11001} (\bibinfo{year}{2011}).

\bibitem[{sli()}]{slipstacking}
\bibinfo{note}{K.~Seiya et.\ al.\ Proc. PAC09 (2009) p1424}.

\bibitem[{\citenamefont{Bocean}(1999)}]{Bocean:1999bp}
\bibinfo{author}{\bibfnamefont{V.}~\bibnamefont{Bocean}}
  (\bibinfo{year}{1999}), \bibinfo{note}{prepared for 6th International
  Workshop on Accelerator Alignment (IWAA 99), Grenoble, France, 18-21 Oct
  1999}.

\bibitem[{\citenamefont{Soler et~al.}(2000)\citenamefont{Soler, Foote, Hoyle,
  and Bocean}}]{Bocean:2000}
\bibinfo{author}{\bibfnamefont{T.}~\bibnamefont{Soler}},
  \bibinfo{author}{\bibfnamefont{R.~H.} \bibnamefont{Foote}},
  \bibinfo{author}{\bibfnamefont{D.}~\bibnamefont{Hoyle}}, \bibnamefont{and}
  \bibinfo{author}{\bibfnamefont{V.}~\bibnamefont{Bocean}},
  \bibinfo{journal}{Geophysical Research Letters}
  \textbf{\bibinfo{volume}{27}}, \bibinfo{pages}{3921} (\bibinfo{year}{2000}).

\bibitem[{\citenamefont{Skaloud and Schwarz}(2000)}]{ref:inertial}
\bibinfo{author}{\bibfnamefont{J.}~\bibnamefont{Skaloud}} \bibnamefont{and}
  \bibinfo{author}{\bibfnamefont{K.}~\bibnamefont{Schwarz}},
  \bibinfo{journal}{Zeitschrift fuer Vermessungswesen (ZfV)}
  (\bibinfo{year}{2000}).

\bibitem[{\citenamefont{Craymer et~al.}(2000)\citenamefont{Craymer, Ferland,
  and Snay}}]{craymer2000realization}
\bibinfo{author}{\bibfnamefont{M.}~\bibnamefont{Craymer}},
  \bibinfo{author}{\bibfnamefont{R.}~\bibnamefont{Ferland}}, \bibnamefont{and}
  \bibinfo{author}{\bibfnamefont{R.}~\bibnamefont{Snay}}, in
  \emph{\bibinfo{booktitle}{Towards an integrated global geodetic observing
  system (IGGOS)}} (\bibinfo{publisher}{Springer}, \bibinfo{year}{2000}), pp.
  \bibinfo{pages}{118--121}.

\bibitem[{\citenamefont{Mortiz}(1980)}]{mortiz1980geodetic}
\bibinfo{author}{\bibfnamefont{H.}~\bibnamefont{Mortiz}},
  \bibinfo{journal}{Bulletin Geodesique} \textbf{\bibinfo{volume}{54}}
  (\bibinfo{year}{1980}).

\bibitem[{ref()}]{ref:commonview}
\bibinfo{note}{David W. Allan and Marc A. Weiss, Proc. 34th Ann. Freq. Control
  Symposium, USAERADCOM, Ft. Monmouth, NJ 07703, May 1980}.

\bibitem[{ste()}]{stefaniaPTTI}
\bibinfo{note}{{S}.~R{\"o}misch, et.\ al.\ Proc.\ 44th PTTI (2012) p99}.

\bibitem[{phi()}]{philPTTI}
\bibinfo{note}{P.~Adamson, et.\ al.\ Proc.\ 44th PTTI (2012) p119}.

\bibitem[{\citenamefont{T\`etreault et~al.}(2005)\citenamefont{T\`etreault,
  Kouba, H\`eroux, and Legree}}]{ref:csrs-ppp}
\bibinfo{author}{\bibfnamefont{P.}~\bibnamefont{T\`etreault}},
  \bibinfo{author}{\bibfnamefont{J.}~\bibnamefont{Kouba}},
  \bibinfo{author}{\bibfnamefont{P.}~\bibnamefont{H\`eroux}}, \bibnamefont{and}
  \bibinfo{author}{\bibfnamefont{P.}~\bibnamefont{Legree}},
  \bibinfo{journal}{Geomatica} \textbf{\bibinfo{volume}{59}},
  \bibinfo{pages}{17} (\bibinfo{year}{2005}).

\bibitem[{\citenamefont{Webb and Zumberge}(1995)}]{webb1995introduction}
\bibinfo{author}{\bibfnamefont{F.}~\bibnamefont{Webb}} \bibnamefont{and}
  \bibinfo{author}{\bibfnamefont{J.}~\bibnamefont{Zumberge}},
  \bibinfo{type}{Tech. Rep.} \bibinfo{number}{JPL D-11088},
  \bibinfo{institution}{Jet Propulsion Lab., Calif. Inst. of Technol.,
  Pasadena} (\bibinfo{year}{1995}).

\bibitem[{\citenamefont{Landau and Lifshitz}(1962)}]{Sagnac}
\bibinfo{author}{\bibfnamefont{L.~D.} \bibnamefont{Landau}} \bibnamefont{and}
  \bibinfo{author}{\bibfnamefont{E.~M.} \bibnamefont{Lifshitz}}, in
  \emph{\bibinfo{booktitle}{{Classical Theory of Fields}}}
  (\bibinfo{publisher}{Pergamon Press}, \bibinfo{year}{1962}), pp.
  \bibinfo{pages}{296--297}.

\bibitem[{\citenamefont{Adamson et~al.}(2011)}]{ref:minosCC2010}
\bibinfo{author}{\bibfnamefont{P.}~\bibnamefont{Adamson}} \bibnamefont{et~al.}
  (\bibinfo{collaboration}{MINOS}), \bibinfo{journal}{Phys.\ Rev.\ Lett.}
  \textbf{\bibinfo{volume}{106}}, \bibinfo{pages}{181801}
  (\bibinfo{year}{2011}).

\end{thebibliography}
